\documentclass[conference]{IEEEtran}

\usepackage[usenames,dvipsnames,svgnames,table]{xcolor}

\usepackage[printonlyused]{acronym}

\acrodef{DPDK}{Data Plane Development Kit}
\acrodef{vRAN}{virtualized Radio Access Network}
\acrodef{ILP}{Integer Linear Program}
\acrodef{VM}{Virtual Machine}
\acrodef{C-RAN}{Cloud Radio Access Network}
\acrodef{RAN}{Radio Access Network}
\acrodef{UPnP}{Universal Plug and Play}
\acrodef{NAT}{Network Address Translation}
\acrodef{DHCP}{Dynamic Host Configuration Protocol}
\acrodef{ISG}{Industry Standards Group}
\acrodef{3GPP}{3rd Generation Partnership Project}
\acrodef{EPC}{Evolved Packet Core}
\acrodef{CPE}{Customer Premises Equipment}
\acrodef{LTE}{Long Term Evolution}
\acrodef{HA}{Hardware Acceleration}
\acrodef{ETSI}{European Telecommunications Standards Institute}
\acrodef{InP}{Infrastructure Provider}
\acrodef{SDN}{Software Defined Networking}
\acrodef{VN}{Virtual Network}
\acrodef{QoE}{Quality-of-Experience}
\acrodef{QoS}{Quality-of-Service}
\acrodef{VNF}{Virtual Network Function}
\acrodef{NFVI}{Network Function Virtualization Infrastructure}
\acrodef{NF}{Network Function}
\acrodef{SN}{Substrate Network}
\acrodef{vCPE}{virtual Customer Premises Equipment}
\acrodef{vEPC}{virtual Evolved Packet Core}
\acrodef{VNE}{Virtual Network Embedding}
\acrodef{SP}{Service Provider}
\acrodef{ARPU}{Average Revenue Per User}
\acrodef{MANO}{Management and Orchestration}
\acrodef{TSP}{Telecommunication Service Provider}
\acrodef{NFV}{Network Function Virtualization}
\acrodef{CAPEX}{Capital Expenses}
\acrodef{OPEX}{Operating Expenses}
\acrodef{DPI}{Deep Packet Inspection}
\acrodef{ONF}{Open Network Foundation}

\ifCLASSINFOpdf
   \usepackage[pdftex]{graphicx}

   \graphicspath{{../pdf/}{figures/}}
   \DeclareGraphicsExtensions{.pdf,.jpeg,.png,.jpg}
\else
\fi
\usepackage[cmex10]{amsmath}
\usepackage{algorithmic}
\usepackage{algorithm}
\usepackage{array}
\newcolumntype{L}[1]{>{\raggedright\let\newline\\\arraybackslash\hspace{0pt}}m{#1}}
\newcolumntype{C}[1]{>{\centering\let\newline\\\arraybackslash\hspace{0pt}}m{#1}}
\newcolumntype{R}[1]{>{\raggedleft\let\newline\\\arraybackslash\hspace{0pt}}m{#1}}

\usepackage{multirow}

\usepackage{url}
\usepackage[table]{xcolor}

\def\b0{{\bf 0}}
\def\bb1{{\bf 1}}
\usepackage[printonlyused]{acronym}

\acrodef{ILP}{Integer Linear Program}
\acrodef{ISG}{Industry Standards Group}
\acrodef{3GPP}{3rd Generation Partnership Project}
\acrodef{EPC}{Evolved Packet Core}
\acrodef{LTE}{Long Term Evolution}
\acrodef{ETSI}{European Telecommunications Standards Institute}
\acrodef{InP}{Infrastructure Provider}
\acrodef{SDN}{Software Defined Networking}
\acrodef{VN}{Virtual Network}
\acrodef{VNF}{Virtual Network Function}
\acrodef{NFVI}{Network Function Virtualization Infrastructure}
\acrodef{NF}{Network Function}
\acrodef{SN}{Substrate Network}
\acrodef{vCPE}{virtual Customer Premises Equipment}
\acrodef{vEPC}{virtual Evolved Packet Core}
\acrodef{VNE}{Virtual Network Embedding}
\acrodef{SP}{Service Provider}
\acrodef{ARPU}{Average Revenue Per User}
\acrodef{MANO}{Management and Orchestration}
\acrodef{TSP}{Telecommunication Service Provider}
\acrodef{NFV}{Network Function Virtualization}
\acrodef{CAPEX}{Capital Expenses}
\acrodef{OPEX}{Operating Expenses}
\acrodef{VEPC}{Virtualized Evolved Packet Core}
\acrodef{VCPE}{Virtualized Customer Premises Equipment}
\acrodef{VRAN}{Virtualized Radio Access Network}

\acrodef{MAS}{Multi-Agent System}
\acrodef{DC}{Data Center}
\acrodef{PAP}{Placement and Assignment Problem}
\acrodef{FPP}{Function Placement Problem}
\acrodef{SPP}{Server Placement Problem}
\acrodef{VDCE}{Virtual Data Center Embedding}
\acrodef{ILP}{Integer Linear Program}
\acrodef{BILP}{Binary Integer Linear Program}
\acrodef{MILP}{Mixed Integer Linear Program}
\acrodef{LARE}{Langragian Relaxation}
\acrodef{VNE}{Virtual Network Embedding}
\acrodef{OTT}{over-the-top}
\acrodef{BPP}{Bin Packing Problem}
\acrodef{MINLP}{Mixed-Integer Nonlinear Programming}
\acrodef{ML}{Machine Learning}
\acrodef{LP}{Linear Program}
\acrodef{CFLP}{Capacitated Facility Location Problem}
\acrodef{ISG}{Industry Standards Group}
\acrodef{ETSI}{European Telecommunications Standards Institute}
\acrodef{PM}{Physical Machine}
\acrodef{VM}{Virtual Machine}
\acrodef{EPC}{Evolved Packet Core}
\acrodef{MCLP}{Maximal Covering Location Problem}
\acrodef{CPRI}{Common Public Radio Interface}
\acrodef{vRAN}{Virtualized Radio Access Network}
\acrodef{LPTRA}{Low Power Transmit and Receive Antenna}
\acrodef{HPTRA}{High Power Transmit and Receive Antenna}
\acrodef{NFV-HRAN}{NFV-based HRAN}
\acrodef{RRH}{Remote Radio Head}
\acrodef{BBP}{Base Band Processing}
\acrodef{BBU}{Base Band Unit}
\acrodef{vBBU}{Virtualized Base Band Unit}
\acrodef{eNodeB}{Evolved Node B}
\acrodef{UE}{User Equipment}
\acrodef{PN}{Physical Network}
\acrodef{VN}{Virtual Network}
\acrodef{HRAN}{Heterogeneous Radio Access Network}
\acrodef{RAN}{Radio Access Network}
\acrodef{CRAN}{Centralized Radio Access Network}
\acrodef{TSP}{Telecommunication Service Provider}
\acrodef{BS}{Base Station}
\acrodef{CSI}{Channel State Information}
\acrodef{QSI}{Queue State Information}
\acrodef{NFV}{Network Function Virtualization}
\acrodef{HPN}{High Power Node}
\acrodef{LPN}{Low Power Node}
\acrodef{RL}{Reinforcement Learning}
\acrodef{LTE}{Long Term Evolution}
\acrodefplural{CAPEX}[CAPEX]{CAPital EXpenses}
\acrodefplural{OPEX}[OPEX]{OPerating EXpenses}

\begin{document}
\title{On the Energy Efficiency Prospects of Network Function Virtualization}

\author{\IEEEauthorblockN{Rashid Mijumbi\IEEEauthorrefmark{2},
Joan Serrat\IEEEauthorrefmark{2},
Juan-Luis Gorricho\IEEEauthorrefmark{2} and
Javier Rubio-Loyola\IEEEauthorrefmark{3}
}
\IEEEauthorblockA{\IEEEauthorrefmark{2}Universitat Polit\`{e}cnica de Catalunya, 08034 Barcelona, Spain}
\IEEEauthorblockA{\IEEEauthorrefmark{3}CINVESTAV, Tamaulipas, Mexico}
%\IEEEauthorblockA{\IEEEauthorrefmark{3}Ghent University $-$ iMinds, B-9050 Gent, Belgium}
%\IEEEauthorblockA{\IEEEauthorrefmark{2}Telecommunications Software and Systems Group, Waterford Institute of Technology, Ireland}
}

\maketitle

\begin{abstract}
\ac{NFV} has recently received significant attention as an innovative way of deploying network services. By decoupling network functions from the physical equipment on which they run, \ac{NFV} has been proposed as passage towards service agility, better time-to-market, and reduced \ac{CAPEX} and \ac{OPEX}. One of the main selling points of \ac{NFV} is its promise for better energy efficiency resulting from consolidation of resources as well as their more dynamic utilization. However, there are currently no studies or implementations which attach values to energy savings that can be expected, which could make it hard for \acp{TSP} to make investment decisions. In this paper, we utilize Bell Labs' GWATT tool to estimate the energy savings that could result from the three main NFV use cases$-$\ac{VEPC}, \ac{VCPE} and \ac{VRAN}. We determine that the part of the mobile network with the highest energy utilization prospects is the \ac{EPC} where virtualization of functions leads to a 22\% reduction in energy consumption and a 32\% enhancement in energy efficiency.
\end{abstract}


\section{Introduction}
\acp{TSP} have been facing an explosion in the data that their networks must support. And this will not stop soon. According to Cisco \cite{Cisco15}, the annual global IP traffic will nearly triple from 2014 to 2019. Specifically, the traffic will pass the zettabyte\footnote{A  zettabyte is equal to 1000 exabytes, which is equal to 1000 billion gigabytes.} threshold by the end of 2016, and reach 2 zettabytes per year by 2019, at which time two-thirds of all this traffic will originate from non-PC devices. With this traffic burst, \acp{TSP} must correspondingly expand and maintain their networks both of which lead to increased \ac{CAPEX} and \ac{OPEX}. Yet, due to competition both among themselves and from services provided over-the-top on their infrastructure, \acp{TSP} cannot respond to these increased costs by increasing subscription fees. This has lead to reductions in \ac{TSP} revenues \cite{nfv}, and forced them to search for ways of deploying and maintaining networks at reduced costs.

\ac{NFV} \cite{Guerzoni12, mano} has been proposed as a viable path towards achieving this goal. By leveraging recent advances in virtualization technology, the main idea of \ac{NFV} is to decouple \acp{NF} from the physical equipment on which they run. This way, the \acp{NF} can be virtualized, and run on industry standard servers and switches which could be located in either data centers or network nodes. The \acp{VNF} may then be relocated and instantiated in different network locations (for example aimed at introduction of a service targeting customers in a given geographical location) without necessarily requiring purchase and installation of new hardware \cite{MijumbiNFV15}.

\begin{figure*}[t]
\begin{minipage}{.50\textwidth}
\centering
\resizebox{.95\textwidth}{!}
{\includegraphics{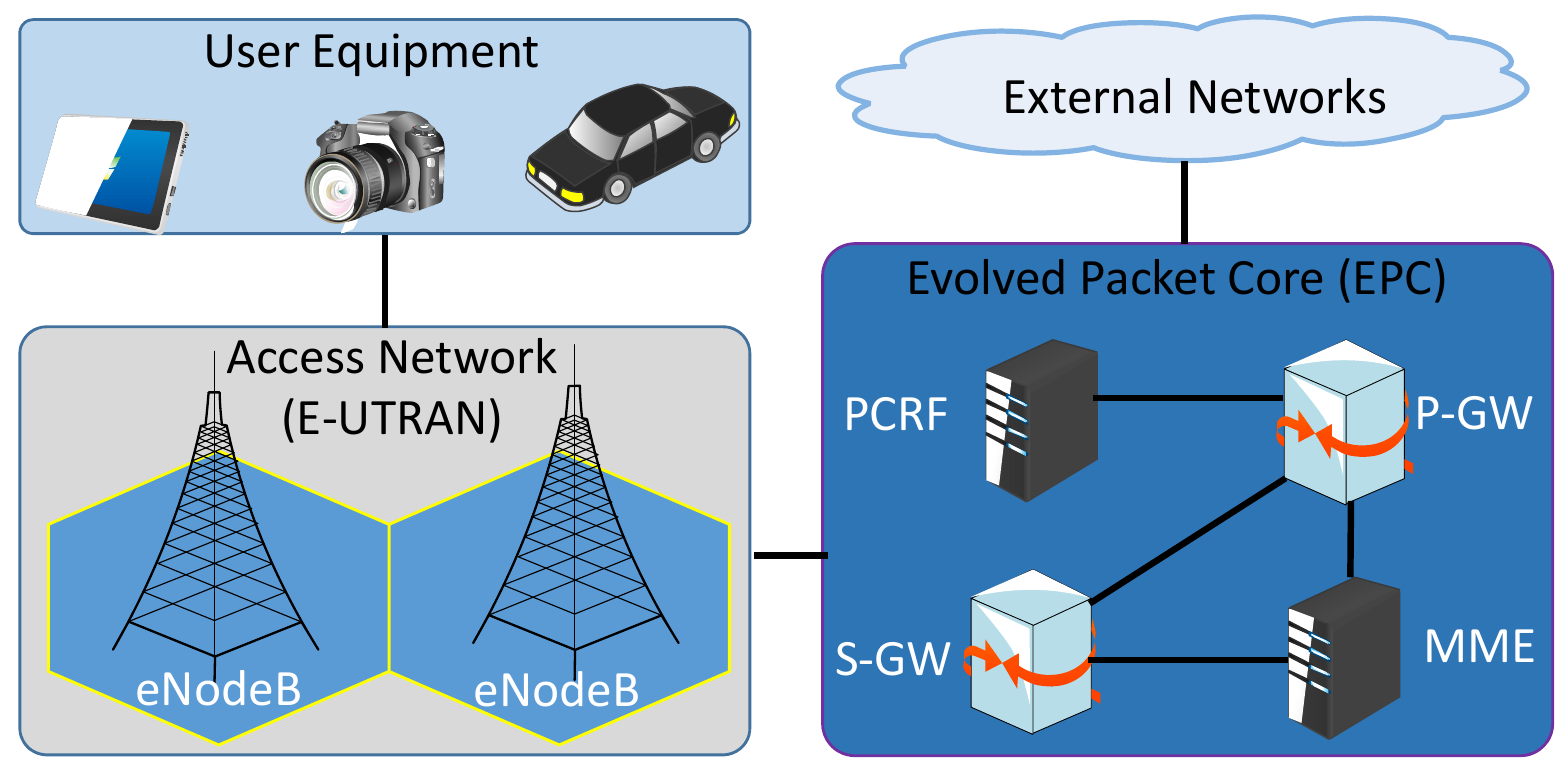}}
  \caption{Traditional \ac{LTE} \ac{EPC}}
  \label{current}
\end{minipage}
\begin{minipage}{.50\textwidth}
\centering
\resizebox{.95\textwidth}{!}
{\includegraphics{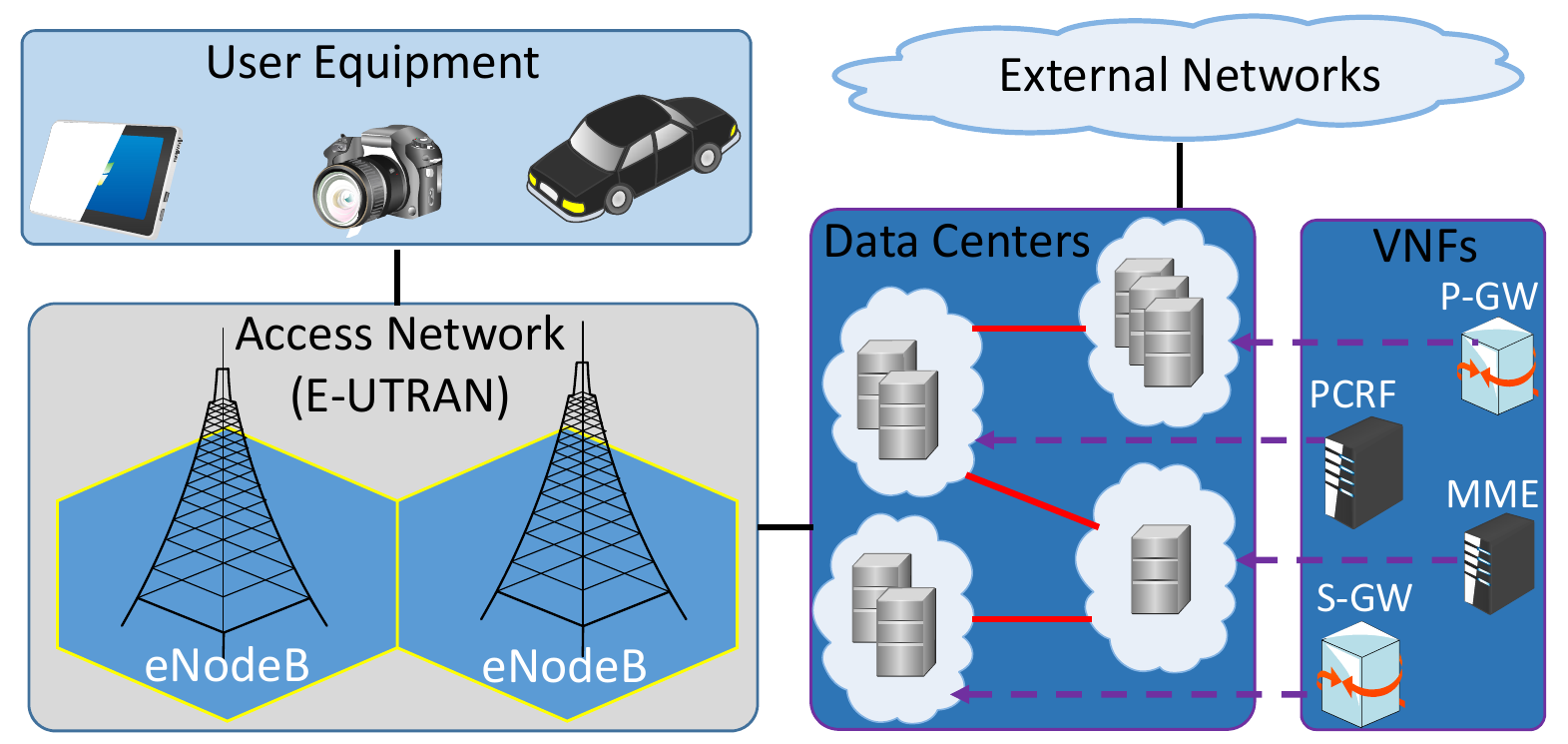}}
  \caption{Possible Virtualized \ac{EPC} Implementation}
  \label{withNFV}
\end{minipage}
\end{figure*}

One of the network parts which is expected to benefit from \ac{NFV} is the \ac{EPC}, which is the core network for \ac{LTE} as specified by 3GPP \cite{3GPP}. In Fig. \ref{current}, we show a basic architecture of \ac{LTE} without \ac{NFV}. The User Equipment (UE) is connected to the \ac{EPC} over the \ac{LTE} access network (E-UTRAN). The evolved NodeB (eNodeB) is the base station for LTE radio. The EPC performs essential functions including subscriber tracking, mobility management and session management. It is made up of four \acp{NF}: Serving Gateway (S-GW), Packet Data Network (PDN) Gateway (P-GW), Mobility Management Entity (MME), and Policy and Charging Rules Function (PCRF). It is also connected to external networks, which may include the IP Multimedia Core Network Subsystem (IMS). In current \ac{EPC} implementations, all its functions are based on proprietary equipment spread across the core network. Therefore, even minor changes to a given function may require a replacement of the equipment. The same applies to cases when the capacity of the equipment has to be changed.

Fig. \ref{withNFV}, shows the same architecture in which the \ac{EPC} is virtualized. In this case, either all functions in the EPC, or only a few of them are transferred to a shared (cloud) infrastructure. Virtualizing the \ac{EPC} could potentially lead to better flexibility and dynamic scaling, and hence allow \acp{TSP} to respond easily and cheaply to changes in market conditions. For example, as represented by the number of servers allocated to each function in Fig. \ref{withNFV}, there might be a need to increase user plane resources without affecting the control plane. In this case, \acp{VNF} such as a virtual MME may scale independently according to their specific resource requirements. In the same way, \acp{VNF} dealing with the data plane might require a different number of resources than those dealing with signaling only. This flexibility is expected to lead to more efficient utilization of resources. Finally, it also allows for easier software upgrades on the \ac{EPC} network functions, which would hence allow for faster launch of innovative services.

%In the proposed \ac{vRAN} , \ac{RAN} functions such as the \ac{BBU} $-$ including \ac{BBP}, Media Access Control (MAC), Radio Link Control (RLC), Packet Data Convergence Protocol (PDCP), Radio Resource Control (RRC), Control and Coordinated Multi-Point (CoMP) transmission and reception $-$ can be dispatched to a \ac{TSP} as an instance of plain software. In Figures \ref{current} and \ref{withNFV}, we show the current implementations of a \ac{RAN} for \ac{LTE} as well as a possible implementation with \ac{NFV}. It can be observed that in a traditional implementation, the \ac{RRH} in each cell is associated with a dedicated \ac{BBU}, which is then connected to the \ac{EPC}. In the \ac{vRAN} scenario, the \ac{BS} servers responsible for the \ac{BBU} functions are transferred to a shared physical infrastructure and virtualized. They are then connected to the \acp{RRH} over front haul links $-$ possibly based on optical fiber. Depending on the number of \acp{BBU} to be virtualized, it is possible to have more than one physical \ac{BBU} servers. The physical \ac{BBU} servers may be located at a data center or distributed across the \ac{TSP}'s \ac{RAN} node locations.

Breaking the bond between functions and physical equipment is expected to lead to several advantages. First, Since the \ac{NF} can be deployed, chained and updated remotely, this promises more flexibility, agility and reduced time-to-market for services. In addition, as it would no longer be necessary to completely change network equipment as network technologies change (to accommodate the data requirements), \ac{CAPEX} could be significantly reduced. Finally, as a result of deploying \ac{NF} on virtualized resources, \ac{NFV} promises the ability to scale resource allocations up and down as traffic demands ebb and flow. This is expected to potentially reduce the number of physical devices operating at any point, and hence reduce \acp{TSP} energy bills. Since energy bills represent more than 10\% of \acp{TSP}' \ac{OPEX}\footnote{This figure can raise up to 45\% for \acp{TSP} in less developed countries\cite{GWATT}.} \cite{GWATT}, reduced energy consumption is one of the strong selling points of NFV. This means that energy efficiency will be critical key performance indicator for NFV.

In this paper, we use Bell Lab's GWATT tool \cite{GWATT} to estimate the energy savings that could result from the three main uses cases (VEPC, VRAN and VCPE \cite{ETSIUseCases}) of NFV. The tool uses state-of-the-art network and traffic models to estimate not only the expected changes in user traffic, but also the resulting effect on network equipment in terms of energy consumption. It also includes possibilities to set different network architectures, including possibilities to select virtualized networks. We believe that an understanding of the actual savings in energy that can be derived from virtualizing different parts of the network is very important in guiding \acp{TSP} to make investment decisions. In particular, at this early stage of NFV where virtualized functions will exist along side those running on specialized hardware, it could be important for \acp{TSP} to determine which part of the network could be virtualized first based on the expected gains.

\indent The rest of this paper is organized as follows: section \ref{related} discusses related work, while the estimation tool used in described in detail in section \ref{tool}. In section \ref{results}, we present and discuss results from utilizing the tool, before concluding the paper in section \ref{concl}.

\section{Related Work}\label{related}

China Mobile recently published \cite{CRAN14} their experiences in deploying a \ac{C-RAN}. One of the tests was performed on their 2G and 3G networks, where it was observed that by centralizing the RAN, power consumption could be reduced by 41\% due to shared air-conditioning. In addition, Shehab et al. \cite{Shehab13} analyzed the technical potential for energy savings associated with shifting U.S. business software to the cloud. The results suggested a substantial potential for energy savings. In fact, the authors noted that if all U.S. business users shifted their email, productivity software, and CRM software to the cloud, the primary energy footprint of these software applications could be reduced by as much as 87\%.

DROP \cite{Bolla2013} is a middleware platform which was originally aimed at creating software routers on top of commodity servers, while hiding the complexity of its modular architecture to control-plane applications and system administrators. It has been recently extended to DROPv2 \cite{Bolla14} which is focused on implementing more efficient power management in SDN. The idea of DROPv2 is to periodically calculate, based on total traffic, the number of forwarding elements that should be put into a standby state, and how the remaining ones should share the available traffic. Both these research activities do not specifically look at virtualized \acp{NF}.

Yathirah et al. \cite{Yathiraj} identified a need to extend Openstack to include a scheduler which is energy-aware. The idea is to use analytics to determine the current status, and use it to schedule Openstack resources. However, the authors do not give any details about solution or the actual resulting energy savings. Instead of relying on generic servers, \cite{Koji14} uses Application Specific Instruction-set Processor (ASIP) to achieve an energy-efficient deployment of a virtualized deep packet inspection (DPI). Based on experiments, the authors argue that they are able to achieve superior energy efficiency compared to DPIs deployed on commodity servers. However, this would be against the general concept of NFV as it runs the functions on specialized servers. In addition, it only considers a specific, light function. Results could be different on a large scale, for more processing intensive functions or chains of functions.

\indent To the best of our knowledge, the work presented in this paper is the first study and discussion of the forecast energy efficiency for any of the NFV use cases. Given the importance of energy efficiency as a key performance indicator for NFV, we hope that the results of this study can help \acp{TSP} make necessary investment decisions with regard to the evolution towards \ac{NFV}. We also hope that this can direct researchers towards proposing energy-aware algorithms to those parts of the network where savings may be highest. It is important to note that the GWATT tool used in this paper has NOT been developed by the authors. Our contribution is in using the tool$-$which is freely available, and aimed at determining energy efficiency in multiple network scenarios$-$to study, simulate, summarize and discuss results specific to chosen NFV use cases.

\section{Bell Labs' GWATT}\label{tool}
 \begin{figure*}[t]
\begin{minipage}{.99\textwidth}
\centering
\resizebox{1.02\textwidth}{!}
{\includegraphics{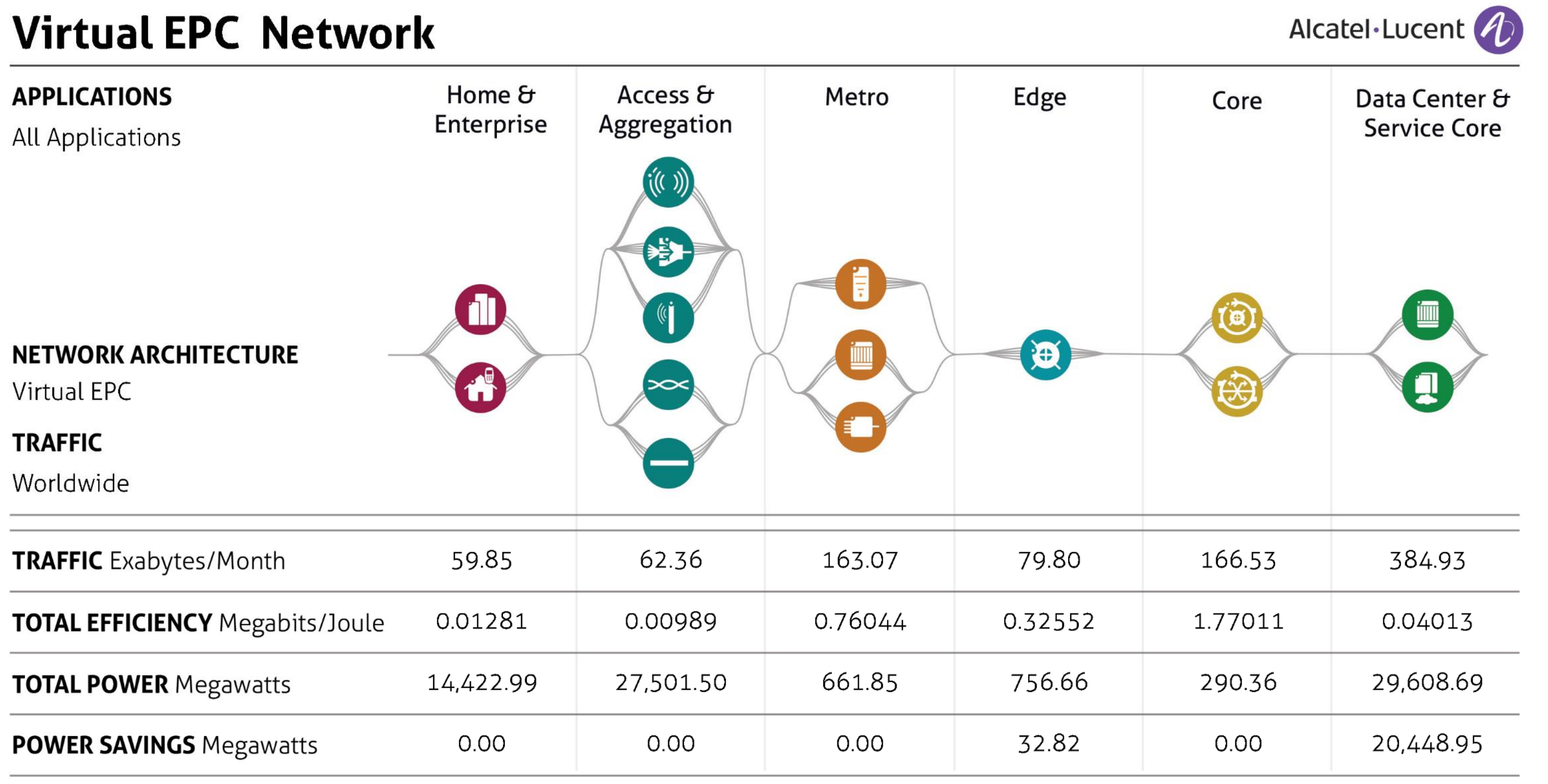}}
  \caption{Network Details from GWATT for Virtualized EPC}
  \label{tool1}
\end{minipage}
\end{figure*}
 \begin{table*}[!htbp]
\caption{Summary of Results for the Baseline Network}
\label{data}
\centering
\includegraphics[width=14.9cm, height=2.5cm]{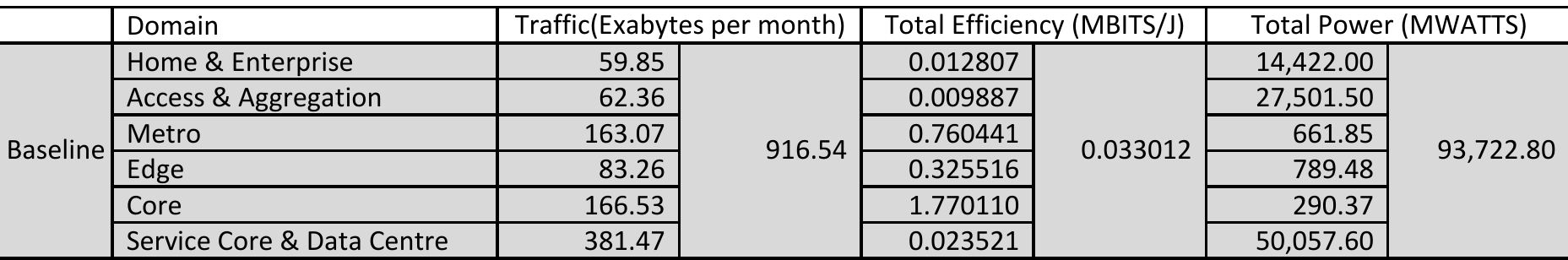}
\end{table*}

 \begin{table*}[t]
\caption{Summary of Results for VEPC, VRAN and VCPE}
\label{data1}
\centering
\includegraphics[width=15.9cm, height=6.0cm]{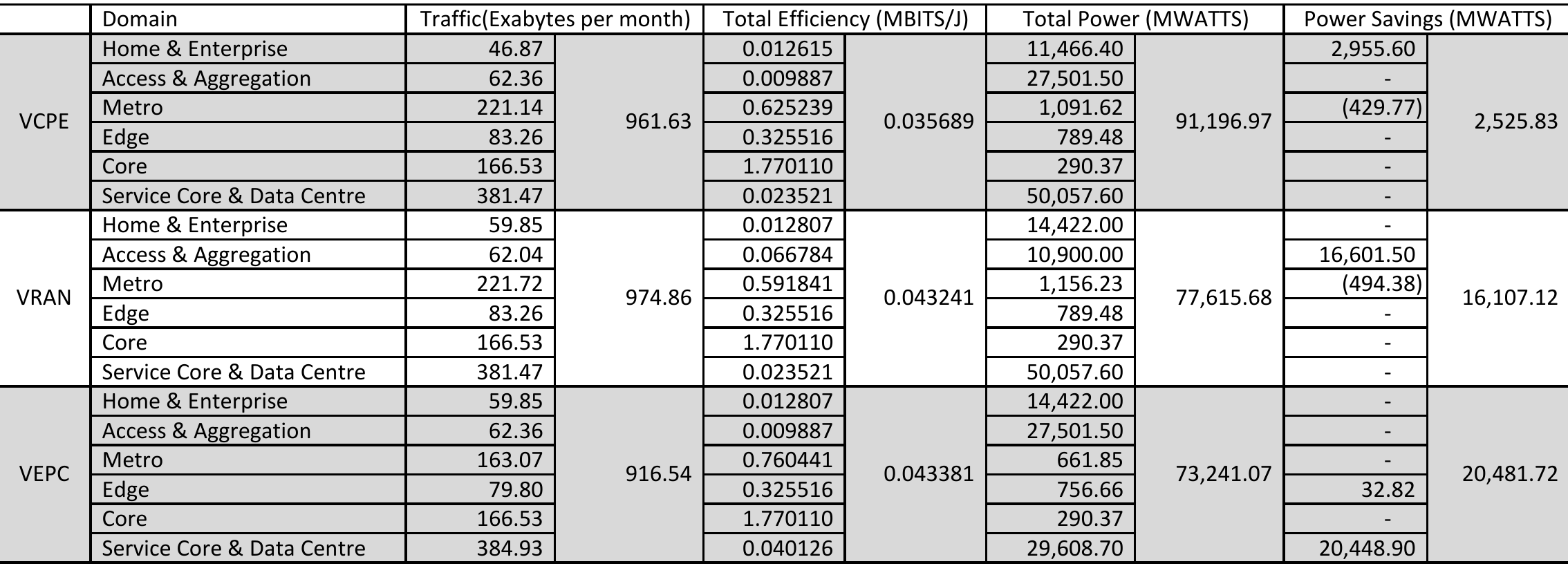}
\end{table*}

Bell Labs' GWATT tool \cite{GWATT} is an important step in determining the possible effect, on energy consumption, of the evolution to \ac{NFV}. Based on some forecast for traffic growth, the tool is able to show the effect of virtualizing different network functions. It also gives a cummulative forecast on energy savings over a five year period. As shown in Fig. \ref{tool1}, the tool divides the network into six domains (Home \& Enterprise, Access \& Aggregation, Metro, Edge, Core and Service Core and Data Centers) which are defined below.

The home \& enterprise is made up of consumer \ac{CPE} in the home, such as DSL routers, set-top boxes and cable modems. It also includes the networking equipment (LAN switches, routers) used within enterprise sites. The access and aggregation is used to connect users to the network using fixed or wireless access, aggregates wireless sites, and provides public switched telephone network (PSTN) service. It includes macro cells each with a tower, antenna, radio and network equipment. It connects multiple wireless cell sites to the mobile operator's network, typically using point-to-point connections. The metro provides aggregation, transport and traffic engineering for residential, business and wireless services. It connects subscribers to a larger service network or the Internet. The edge serves as the primary entry point to service provider core networks and runs service routing, intelligence and signaling functions. The core provides high performance IP routing and optical transport for the service provider's backbone network and internet connection. It includes optical transport systems which can support multiple 10 Gbps channels on each fiber and provide the high speed optical backbone to transport IP core traffic. Finally, the service core and data center domain consists of telecommunication operational systems and the data center equipment to host customer content and run customer applications. The service core consists of multiple telecom service delivery platforms and operational IT systems required to deliver telecommunications services and run the business.

Each of the six network domains can be edited to select different network models and technologies and hence analyze its energy impact. As shown in the top left corner of Fig. \ref{tool1}, the figure is from an analysis of the energy consumption of a network in which the EPC is virtualized. The results from the analysis are shown at the bottom of the figure. To discuss these results, we start by noting that GWATT is based on two main models:
\begin{figure*}[!ht]
\begin{minipage}{.33\textwidth}
\centering
\resizebox{.95\textwidth}{!}
{\includegraphics{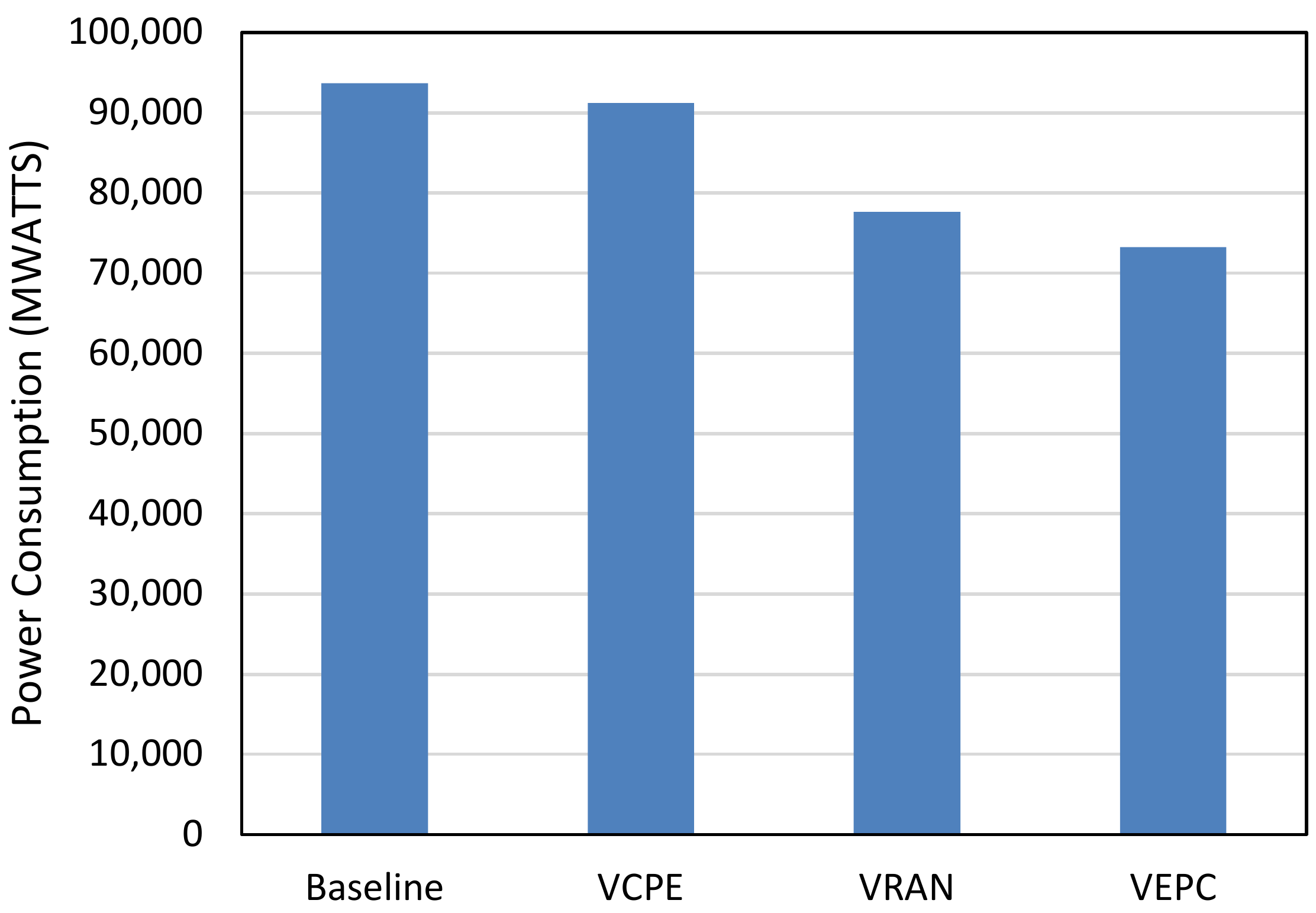}}
  \caption{Total Power}
  \label{totalpower}
\end{minipage}
\begin{minipage}{.33\textwidth}
\centering
\resizebox{.95\textwidth}{!}
{\includegraphics{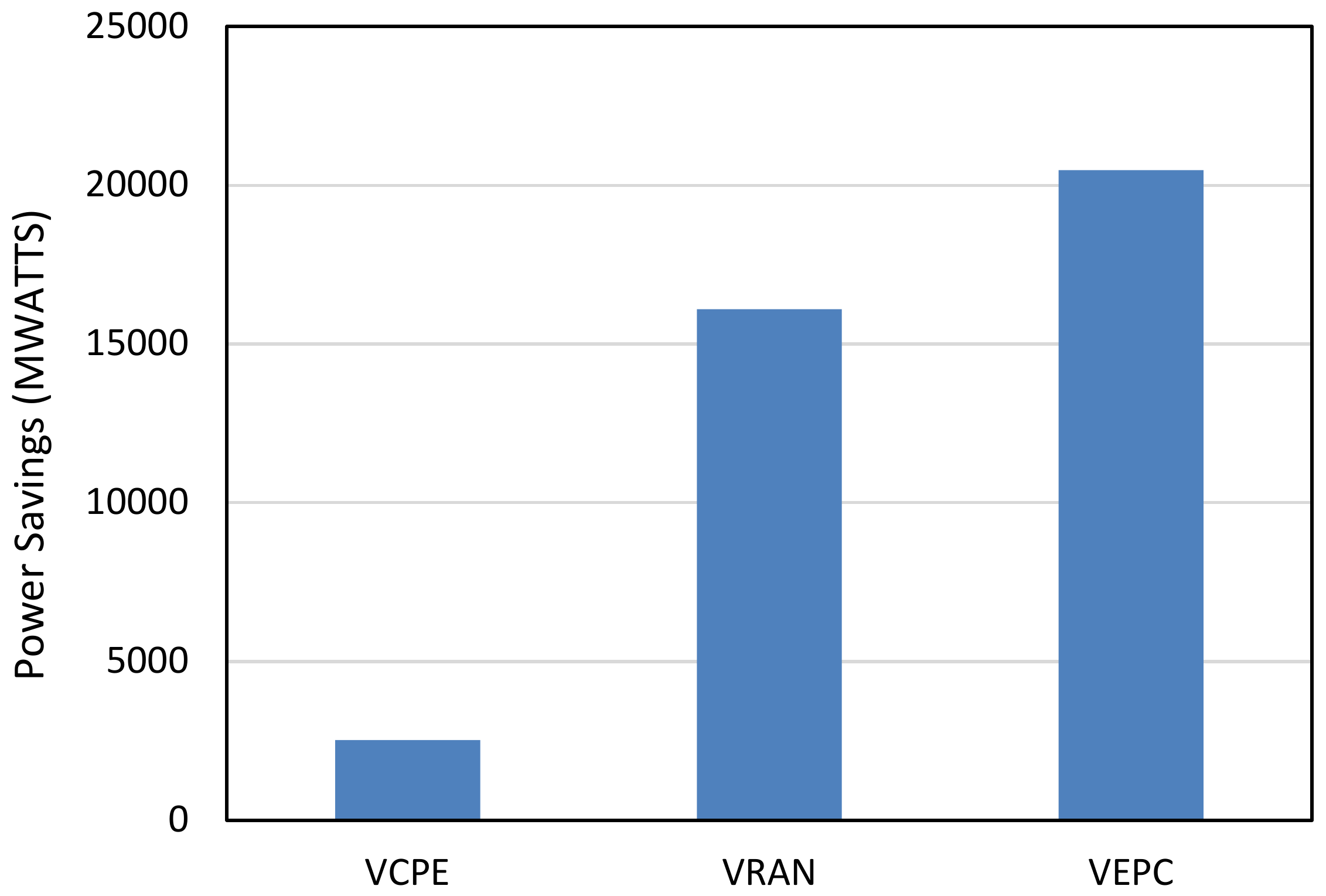}}
  \caption{Power Savings}
  \label{powersavings}
\end{minipage}
\begin{minipage}{.33\textwidth}
\centering
\resizebox{.95\textwidth}{!}
{\includegraphics{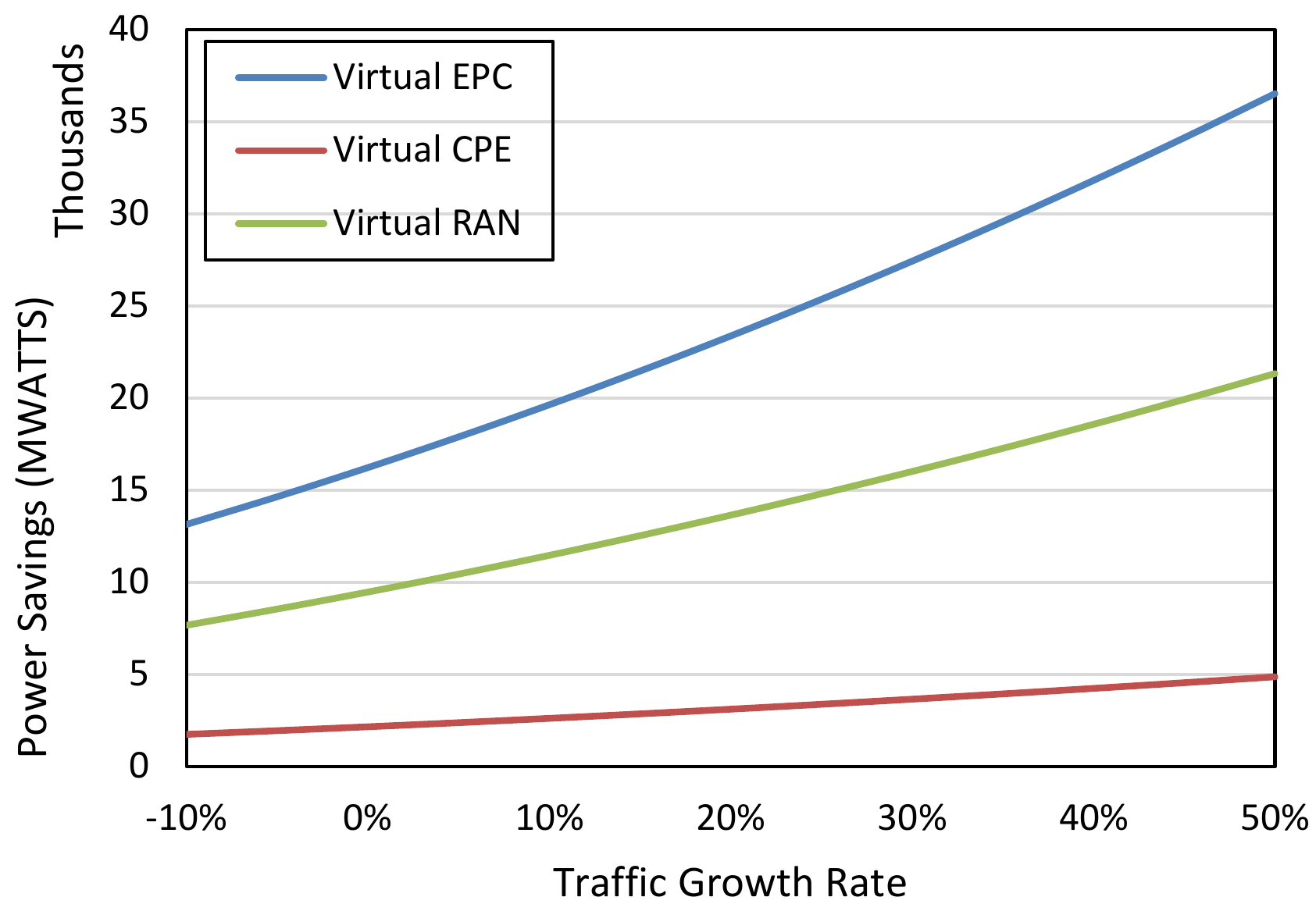}}
  \caption{Variation of Power Savings}
  \label{powersavingseffect}
\end{minipage}
\end{figure*}

\begin{enumerate}
\item A traffic data model. For each domain of the network, the tool uses a traffic volume expressed in exabytes/month. The traffic volume is based on extrapolated traffic projections performed by Bell Labs and based on real data. For example, as shown in Figure \ref{tool1}, when the EPC is virtualized, the traffic associated to the network core domain is $166.53$ exabytes/month, while that associated with the data center and service core is $384.93$ exabytes/month.\\
\item A network element efficiency model. An efficiency expressed in megabits/joules is associated with each domain. For instance, after virtualizing the EPC as shown in Figure \ref{tool1}, it can be observed that the power efficiency for the core network domain is $1.77011$ megabits/joules while that for the data center and service core domain is $0.04013$ megabits/joules. We also note that with this efficiency, the total power consumed by all network elements in each of these domains is $290.36$ Megawatts and $29,608.69$ Megawatts respectively.
\end{enumerate}

Based on the traffic and efficiency data above, and taking into consideration the underlying network model, GWATT computes the total end-to-end network power (expressed in mega watts), energy consumption (in gigajoules). In addition, by comparing all these values with those from a baseline network\footnote{A baseline network is one where all functions are run in physical equipment, using the tool's default technologies and settings.}, the tool is able to determine the forecast savings in energy by changing a given part of the end-to-end network. It can be observed that by virtualizing the EPC, $32.82$ Megawatts of power are saved at the edge while $20,442.95$ Megawatts are saved in the data center and service core. These traffic, consumption and efficiency values for the baseline network are shown in Table \ref{data}.

In order to use the tool, some choices need to be made\footnote{Unless stated otherwise, the results presented in this paper are based on GWATT's default settings.}. These settings relate to the network size, traffic growth rate, number of applications and network architecture. The default network size worldwide, but this can be set to one of North America, Small Country, Large Country etc. The tool also allows one to set the rate at which traffic grows in the selected network size as a percentage. The traffic applications setting may be used to select one or more applications of the available application which include online gaming, file sharing, video, web, business video, business file sharing and business web. The default setting utilizes all the applications, which represents 100\% of the set traffic. Finally, it is possible to select a particular network architecture. While the tool includes many architectures, the focus of this paper is on the virtualization of the CPE, RAN and EPC. The VEPC virtualizes mobile network functions by implementing a cloud-based EPC architecture that uses resources more efficiently and improves agility and scale. The VRAN virtualizes and centralizes the consolidation of eNodeB Baseband Units (BBU) from multiple cell sites to improve scale and efficiency. Finally, the VCPE virtualizes the residential gateway of the CPE to reduce complexity, energy consumption and deployment cost (CAPEX).

 \begin{figure*}[t]
\begin{minipage}{.33\textwidth}
\centering
\resizebox{.95\textwidth}{!}
{\includegraphics{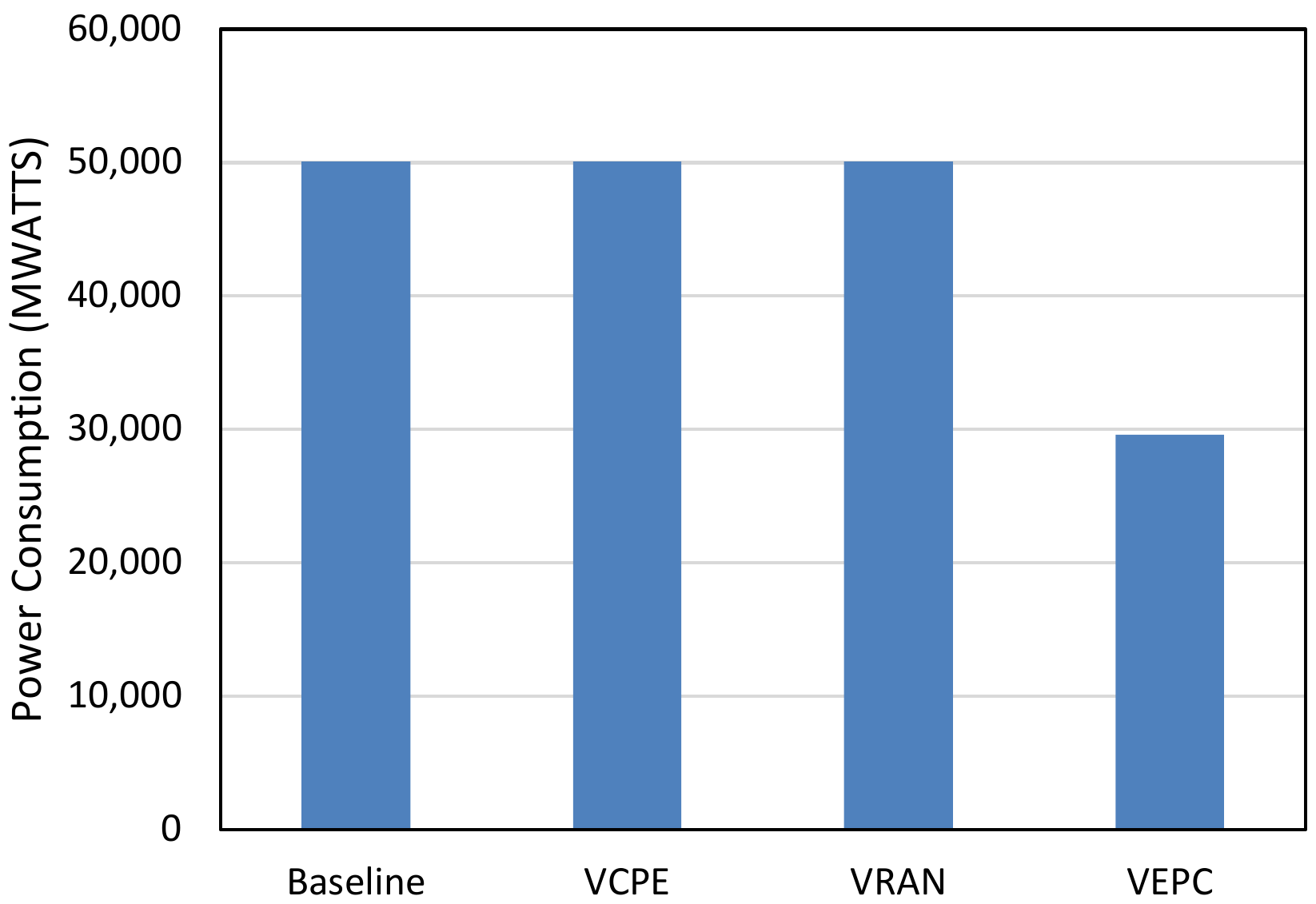}}
  \caption{Power Consumption of Service Core}
  \label{servicecore}
\end{minipage}
\begin{minipage}{.33\textwidth}
\centering
\resizebox{.95\textwidth}{!}
{\includegraphics{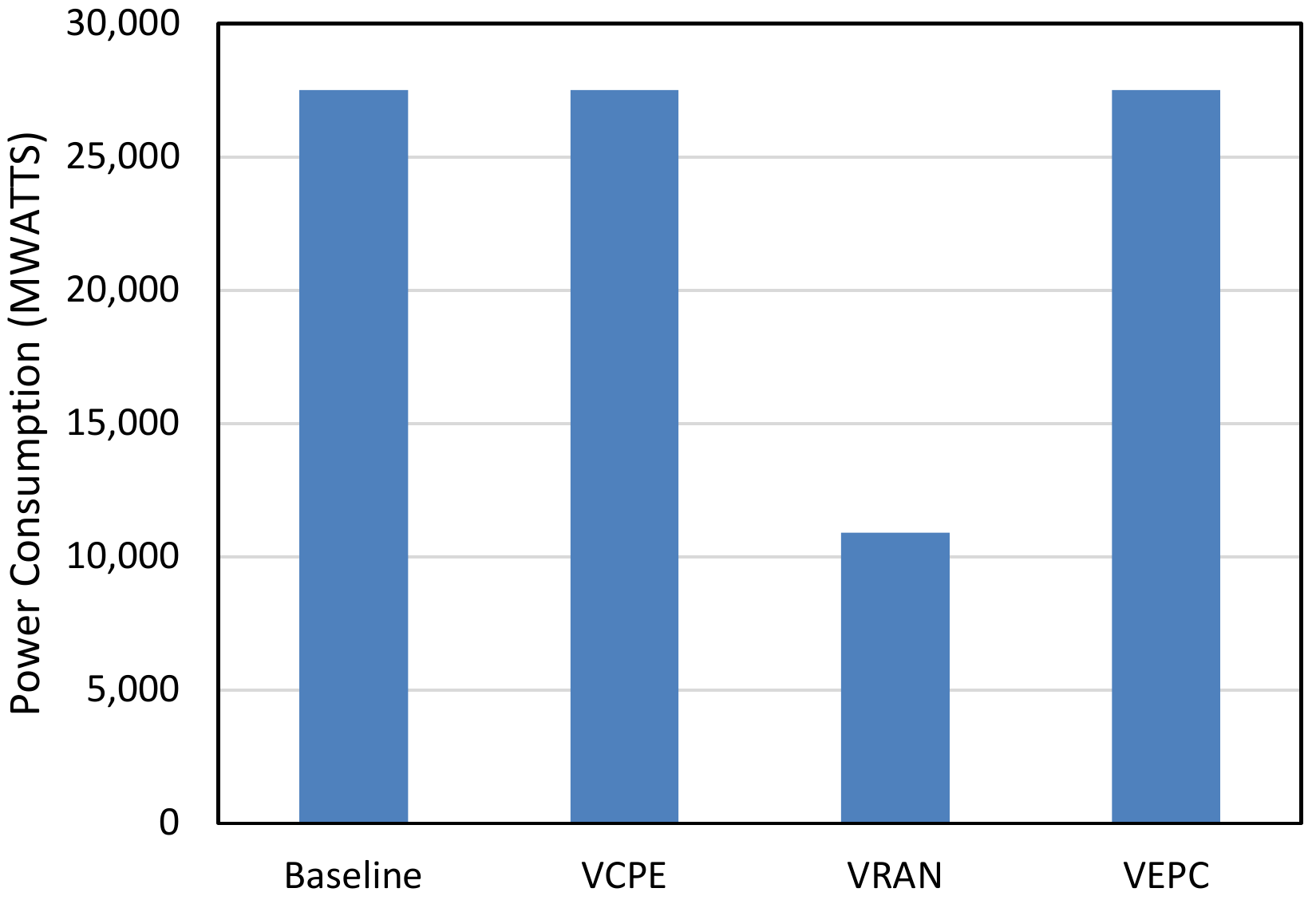}}
  \caption{Power Consumption of Access Network}
  \label{access}
\end{minipage}
\begin{minipage}{.33\textwidth}
\centering
\resizebox{.95\textwidth}{!}
{\includegraphics{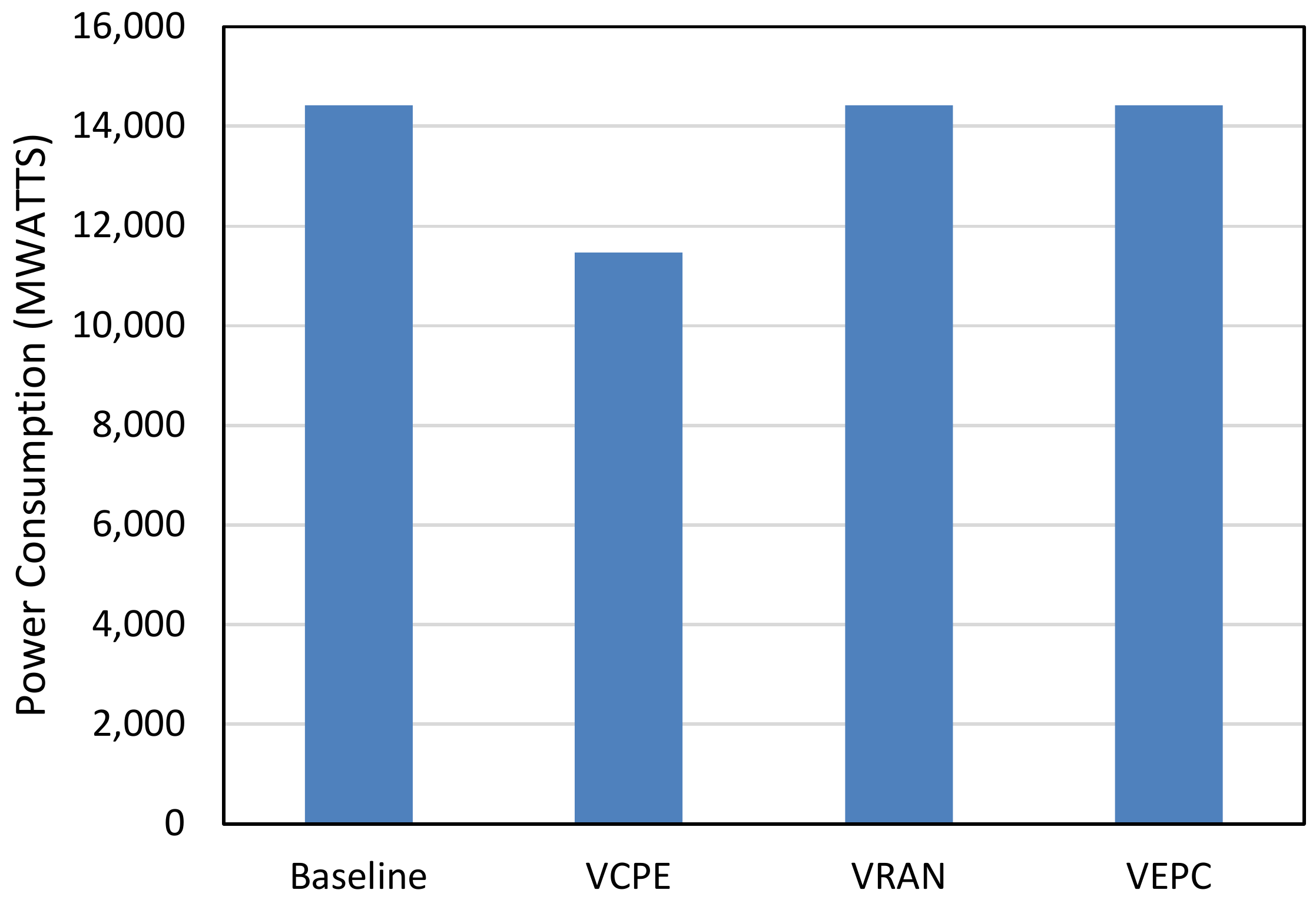}}
  \caption{Power Consumption of Home Network}
  \label{cpe}
\end{minipage}
\end{figure*}

 \begin{figure*}[t]
\begin{minipage}{.30\textwidth}
\centering
\resizebox{.95\textwidth}{!}
{\includegraphics{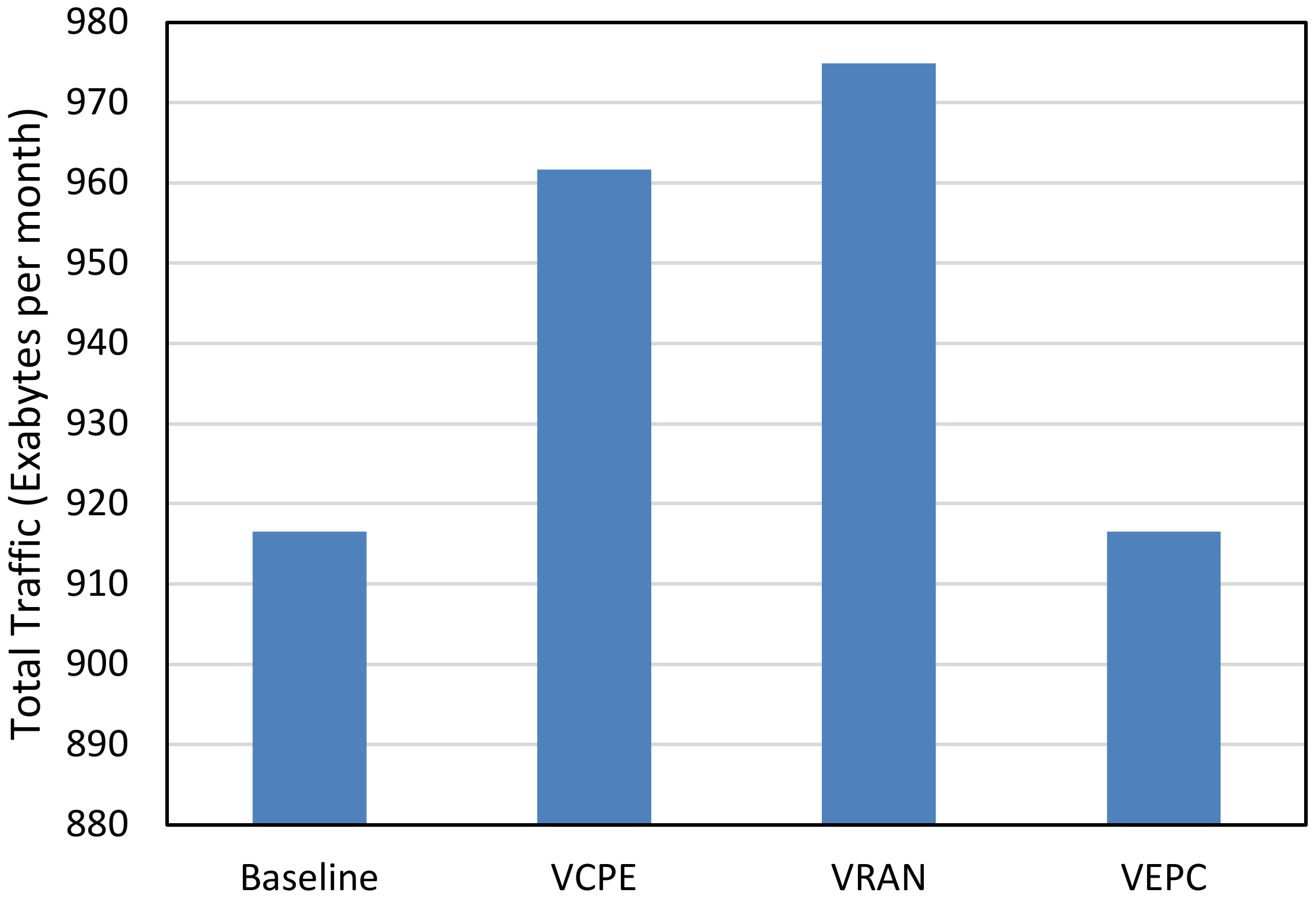}}
  \caption{Total Traffic}
  \label{totaltraffic}
\end{minipage}
\begin{minipage}{.40\textwidth}
\centering
\resizebox{.95\textwidth}{!}
{\includegraphics{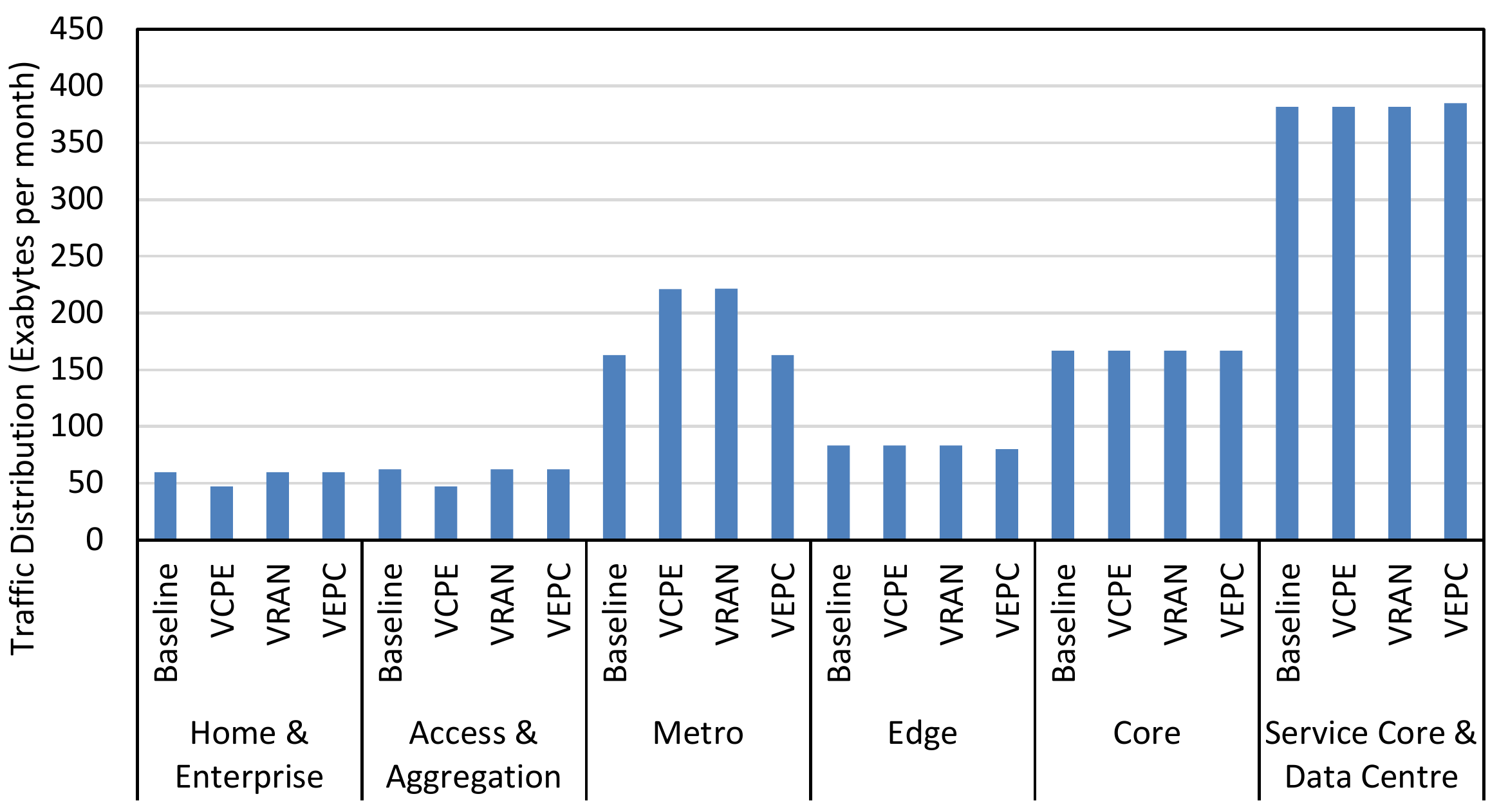}}
  \caption{Traffic Distribution}
  \label{distributed}
\end{minipage}
\begin{minipage}{.30\textwidth}
\centering
\resizebox{.95\textwidth}{!}
{\includegraphics{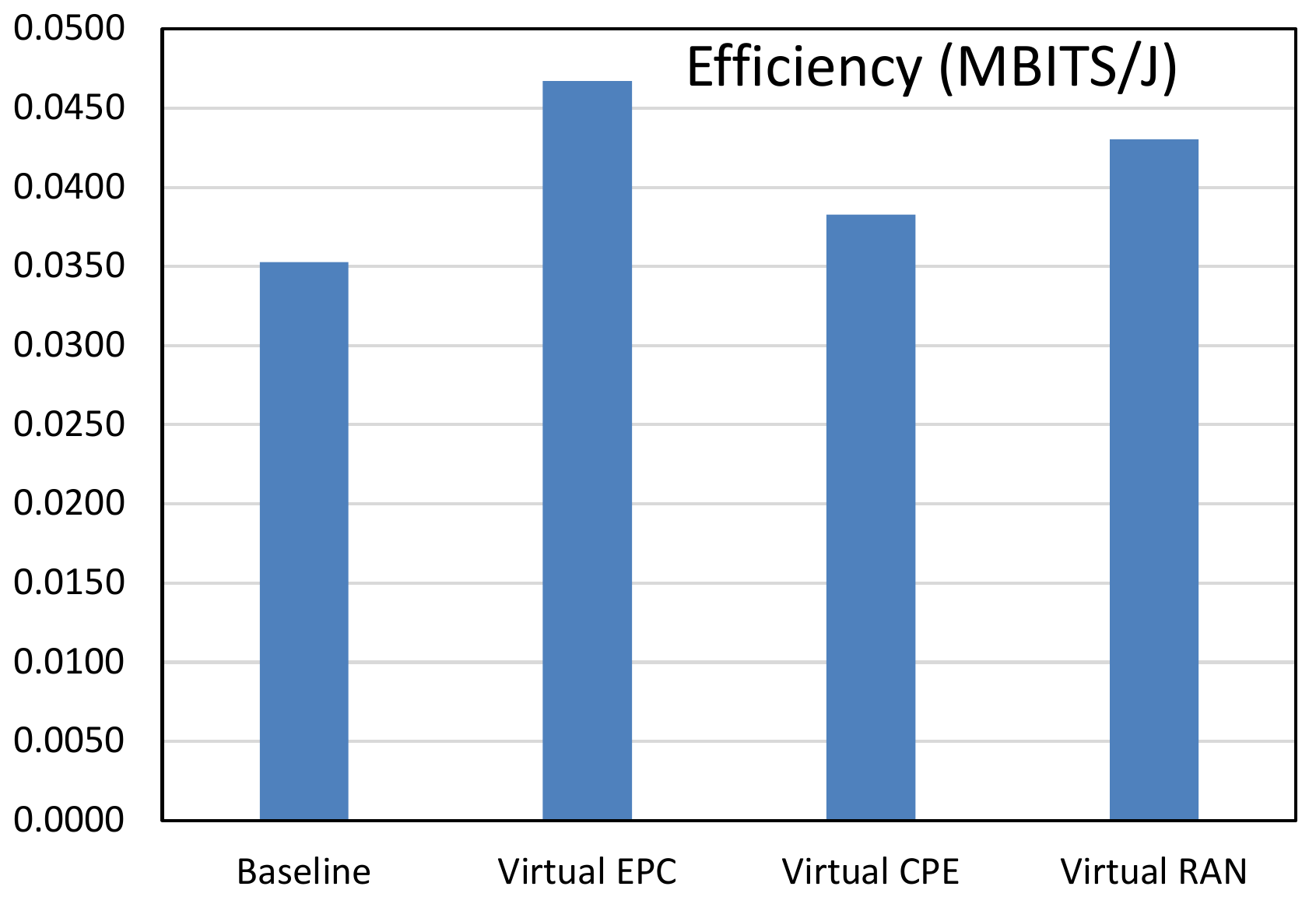}}
  \caption{Efficiency}
  \label{efficiency}
\end{minipage}
\end{figure*}

\section{Results}\label{results}

The results for the three NFV use cases are summarized in Table \ref{data1}. The power savings are determined in comparison with the results from the baseline network in \ref{data}. These results, including others with different setting are presented in graphical form in Figs. \ref{totalpower} - \ref{efficiency}.

From Fig. \ref{totalpower} and \ref{powersavings}, it can be observed virtualization of network functions leads to a reduced power consumption. However, it can be noted that VEPC achieves the highest reduction in power consumption, followed by VRAN and then VCPE. The reason for superiority of virtualizing the EPC could be due to the higher traffic that goes through the EPC domain as shown in Table \ref{data}. This means that there are many network elements in the core, which presents more possibility to reduce energy not only by consolidating these network elements, but by also reducing the required capacity through the better resource management presented by virtualization. In fact, this can be further proved by looking at Fig. \ref{powersavingseffect} which shows how the power savings change if the rate of traffic growth is increased. It can be noted while the VEPC saves 21\% of power even when traffic is reducing, the savings increase to about 23\% in conditions of traffic growing at 50\%.

In Figs. \ref{servicecore}, \ref{access} and \ref{cpe}, we show how virtualizing a given network domain affects the energy consumption of the other domains. As expected, it can be observed that each of the three use cases considered have the strongest effect in terms of power savings on the respective domain which has been virtualized.

In Fig. \ref{totaltraffic}, we show the total network traffic carried in each use case, which is broken up across all the contributing network domains in Fig. \ref{distributed}. As earlier noted, it is evident that the service core and data center domain has the highest traffic in all cases. However, it is interesting to note that while this traffic is almost constant across all the other network domains for the 4 cases evaluated (including the baseline), it varies considerably in the metro domain. In fact, the total traffic traffic in Fig. \ref{totaltraffic} takes on a similar profile to the traffic in the metro network. This could mean that the virtualization of any of the considered use cases could benefit more from virtualization of the metro network. The idea is that if the traffic increases, there is potential to benefit for improved resource management through virtualization of resources.

Finally, \ref{efficiency} shows the total efficiency resulting from virtualizing of the considered domains. Just like the savings in power consumption, and probably for the same reasons, the EPC benefits more from virtualization than both the RAN and CPE. The fact that the CPE lags in terms of energy efficiency and/or power savings in all these results is not so surprising. CPEs normally consume little energy and the advantages from virtualizing the CPE could be much more on the side of CAPEX (as the boxed would become cheaper on a large scale). The OPEX savings will mainly be due to ability to make remote configurations as well as upgrades.

However, while the results presented in this paper show promising values on energy savings, a lot of work will likely need to be done to achieve them in practice. The fact that NFV makes data centers an indispensable part of telecommunication networks raises questions on the actual value of savings in energy consumption (if any) that will be achieved. According to an analysis in the SMARTer 2020 report from GeSI \cite{GESI}, the total electricity requirements of the cloud (including data centers and networks, but not devices) in 2011 was 684 billion kWh. The cloud, if it were a country, would rank 6th in the world in terms of its energy demand, and yet this demand is expected to increase by 63\% by 2020 \cite{CLICKCLEAN}. While some progress on energy efficient cloud computing has been made, the fast growing energy needs of data centers continue to receive a lot of attention \cite{USDCS, BeloglazovBLZ11}. Therefore, there is an urgent need to evaluate whether NFV will be able to meet these energy saving expectations, or whether$-$like the NFs$-$the energy consumption will just be transferred to the cloud. 

\section{Conclusion}\label{concl}

In this paper, we have attached values to the energy saving expectations of some NFV use cases. Using Bell Lab's GWATT tool, we have determined that by virtualizing the EPC, RAN and CPE we can reduce network power consumption by 22\%, 17\% and 3\% respectively in comparison to a baseline worldwide network consumption of 93,722.80 Megawatts. We have also shown that this saving in energy consumption could even be higher in cases of increasing traffic across the network. Finally, we have noted that virtualizing each of these network domains does not significantly affect the traffic passing through the domain, instead affecting the traffic through the metro network. For this reason, its important to consider the virtualization of the metro network as a necessary aspect of virtualizing other network domains.

However, while GWATT is an important step in attaching numbers to the energy savings expected from NFV, it can still be improved. In particular, it currently does not have a detailed technical documentation. In addition, Cisco's visual
networking index \cite{Cisco15} forecasts that annual global IP traffic will reach 1000 exabytes in 2016. Based on this, the (monthly) traffic values in Tables \ref{data} and \ref{data1} seem to be so high, yet the tool does not provide a possibility to know how these values are derived. Without a doubt, the energy efficiency of cloud based NFs will continue to receive attention. NFV will put InPs under even more pressure to manage energy consumption not to only to cut down energy expenses, but also to meet legal, regulatory and environmental requirements. Therefore, approaches for placing network functions \cite{path, aims} as well as those that dynamically manage the network resources \cite{rl, neurofuzzy, sdn, neural} must be geared towards energy efficiency just as they are focused on efficient resource utilization.

%Therefore, we expect that the energy efficiency of cloud based NFs will continue to receive attention. NFV will put InPs under even more pressure to manage energy consumption \cite{Bolla14} not to only to cut down energy expenses, but also to meet regulatory and environmental standards. Topics with regard to energy efficient hardware which could allow reductions in CPU speeds and partially turning off some hardware components, more energy-aware function placement, scheduling and chaining algorithms, will be important. An example could be to track the cheapest prices for energy costs and adapt the network topology and/or operating parameters to minimize the cost of running the network \cite{Cui14}. However, all these should be carefully considered to ensure that there is a balance in the trade-off between energy efficiency and function performance or service level agreements.
%
%However, while the tool is an important step in attaching numbers to the energy savings expected from NFV, it can still be improved. In particular, it does not yet have a detailed technical documentation. For example, Cisco's visual networking index \cite{Cisco15} forecasts that annual global IP traffic will reach 1000 exabytes in 2016. Based on this, the (monthly, 2015) traffic values in Table \ref{gwatttable} seem to be so high, yet it is currently not possible to know how these values are derived. 

\section*{Acknowledgment}
This work is partly funded by FLAMINGO, a Network of Excellence project (318488) supported by the European Commission under its Seventh Framework Programme, and project TEC2012-38574-C02-02 from Ministerio de Economia y Competitividad.

\bibliographystyle{IEEEtran}
\bibliography{IEEEabrv,cnsmnfvbiblio}

\end{document}